\documentstyle[epsfig]{aipproc}
\def\gtaprx {\lower .1ex\hbox{\rlap{\raise .6ex\hbox{\hskip .3ex
	{\ifmmode{\scriptscriptstyle >}\else
		{$\scriptscriptstyle >$}\fi}}}
	\kern -.4ex{\ifmmode{\scriptscriptstyle \sim}\else
		{$\scriptscriptstyle\sim$}\fi}}}
\def\ltaprx {\lower .1ex\hbox{\rlap{\raise .6ex\hbox{\hskip .3ex
	{\ifmmode{\scriptscriptstyle <}\else
		{$\scriptscriptstyle <$}\fi}}}
	\kern -.4ex{\ifmmode{\scriptscriptstyle \sim}\else
		{$\scriptscriptstyle\sim$}\fi}}}
\def \sun {$_{\scriptscriptstyle \odot}$}
\def \sune {$_{\scriptscriptstyle \odot}$}

\begin{document}
\title{Gamma-Ray Bursts: The Central Engine}

\author{S. E. Woosley} 
\address{Astronomy Department, UCSC\thanks{This
research was supported by the National Science Foundation,
(AST-97-31569) and by the NASA Theory Program (NAG5-8128.)}}

\maketitle

\begin{abstract}
A variety of arguments suggest that the most common form of gamma-ray
bursts (GRBs), those longer than a few seconds, involve the formation
of black holes in supernova-like events. Two kinds of ``collapsar''
models are discussed, those in which the black hole forms promptly - a
second or so after iron core collapse - and those in which formation
occurs later, following ``fallback'' over a period of minutes to
hours. In most cases, extraction of energy from a rapidly accreting
disk (and a rapidly rotating black hole) is achieved by
magnetohydrodynamical processes, although neutrino-powered models
remain viable in cases where the accretion rate is $\gtaprx 0.05$
M\sun \ s $^{-1}$. GRBs are but one observable phenomenon accompanying
black hole birth and other possibilities are discussed, some of which
(long, faint GRBs and soft x-ray transients) may await discovery. Since
they all involve black holes of similar mass accreting one to several
M\sun, collapsars have a nearly standard total energy, around
10$^{52}$ erg, but both the fraction of that energy ejected as highly
relativistic matter and the distribution of that energy with angle can
be highly variable.  An explanation is presented why inferred GRB
luminosity might correlate inversely with time scales and arguments
are given against the production of ordinary GRBs by supergiant stars.
\end{abstract}

\section*{General GRB Model Requirements}

Given the locations afforded by x-ray and optical afterglows,
redshifts have now been determined for approximately 10 GRBs so that
we have at least a small sampling of GRB energies
\cite{Galama99}. They are by no means standard candles.  Even
discounting the unusual case of GRB 980425 ($8 \times 10^{47}$ erg),
energies range from about $5 \times 10^{51}$ erg (GRB 980613) to $2
\times 10^{54}$ erg (GRB 990123). In addition, GRB time profiles and
spectra are very diverse and separate into at least two classes - the
``short-hard'' bursts (with average duration 0.3 s) and the
``long-soft'' bursts (average duration 20 s). The first challenge any
model builder must confront is deciding just which GRBs, and which
features, he or she is attempting to model since it is increasingly
doubtful that all GRBs are to be explained in the same way. Moreover,
in all of today's models, the gamma-rays observed from a cosmological
GRB are produced far from the site where the energy is initially
liberated - presumably conveyed there by relativistic outflow or
jets. How much of what we see in GRB reflects the central engine and
how much the environment where the outflow dissipates its energy? So
our first step is to define the problem we are attempting to address.

Even after the dramatic progress of the last two years, few definitive
statements can be made about GRBs without provoking
controversy. Still, in 1999, most people feel that the following are
facets of a common GRB that the central engine must provide:

\vskip 0.15 in
\noindent
1) Highly relativistic outflow - $\Gamma \gtaprx 100$, possibly highly
collimated.

\vskip 0.15 in

\noindent
2) An event rate that, at the BATSE threshold and in the BATSE energy
range, is about 1/day. Beaming, of course, raises this number
appreciably.

\vskip 0.15 in

\noindent
3) Total energy in relativistic ejecta $\sim 10^{53} - 10^{54} \,
\epsilon_{\gamma}^{-1} \, f_{\Omega}$ erg where $\epsilon_{\gamma}$ is
the efficiency for turning relativistic outflow into gamma-rays ($\sim
10$\%?), and $f_{\Omega}$ is the fraction of the sky into which that
part of the flow having sufficiently high $\Gamma$ ($\gtaprx$100) is
collimated ($\sim$1\%?).  For reasonable values of these parameters,
the total energy required for a common GRB is 10$^{52}$ erg. Fainter
GRBs can result from the same 10$^{52}$ erg event if the efficiency
for producing relativistic matter is reduced (e.g., GRB 980425); brighter 
ones if the collimation is tighter.

\vskip 0.15 in

\noindent
4) A duration of relativistic flow in our direction no longer than the
duration of the GRB. This constraint is highly restrictive for the
short (0.3 s) bursts and may imply multiple models. For GRB models
produced by internal shocks, the flow may additionally need to last
{\sl as long as} the GRB ({\sl modulo} the relativistic time
dilation). This makes a natural time scale $\sim 10$ s attractive.

\vskip 0.15 in

\noindent
5) In the case of long bursts, association with star forming regions in 
galaxies and, in perhaps three cases, with supernovae of Type I.

\vskip 0.2 in

The near coincidence of 10$^{52}$ erg with the energy released in the
gravitational collapse of a stellar mass object to a neutron star (or,
equivalently, the accretion disk of a black hole), has long suggested a
link between GRBs and neutron star or black hole formation, a
connection championed by Paczynski before cosmological models became
fashionable. Viable models separate into three categories (Table 1),
where $\epsilon_{\rm MHD}$ is the unknown efficiency for
magnetohydrodynamical processes to convert either gravitational
accretion energy at the last stable orbit ($\sim 0.1 \dot M c^2$) or
neutron star rotational energy into relativistic outflow. Those using
black hole accretion \cite{Eberl99,Mac99,Ross99,Mes99} typically
employ 1 - 10\% for $\epsilon_{\rm MHD}$; pulsar advocates
\cite{Usov94,Wheel99} need approximately 100\%.

\begin{table}[h!]
\caption{Gravitational collapse models for GRBs}
\label{table1}
\begin{tabular}{ccccccc}
Model & Energy & Mass      & Possible  & Jet  & $^{56}$Ni & Beaming \\
      & Source & Reservoir &   Energy  & From & /SN       &         \\
\\
\tableline
\\
n$^*$+n$^*$,BH  & BH accretion  & 0.01 - 0.5 M\sune
& 10$^{50}$ & $\nu \bar \nu$ & No & $\sim$10\%?\\
                &                &              
& 10$^{53} \epsilon_{\rm MHD}$  & MHD &  &    \\
\\
collapsar  & BH accretion  & 1 - 5 M\sune & 10$^{52}$ & $\nu \bar \nu$ & Yes & 0.1\% - 10\% \\
           &               &  & 10$^{54} \epsilon_{\rm MHD}$  & MHD &  &    \\
\\
pulsar &n$^*$ rotation &  & 10$^{52} \epsilon_{\rm MHD}$ & MHD & ? & $\sim$1\%? \\
\\
\tableline
\end{tabular}
\end{table}

The collapsar model is incapable of producing relativistic jets of
total duration less than a few seconds (hence short hard bursts are
difficult - impossible unless the beam orientation wanders). Merging
neutron stars and black holes, on the other hand, can produce short
bursts if the disk viscosity is high (i.e., $\alpha \sim 0.1$), but
cannot, with the same disk viscosity, produce long bursts. Merging
neutron stars also lack the massive disks that help to focus the
outflow in collapsars and it may be more difficult for them to emit
highly collimated jets. Hence their ``equivalent isotropic energies''
may be smaller (unless MHD collimation dominates). It seems more
natural to associate the merging compact objects with short hard
bursts, but this conjecture presently lacks any observational basis.
Hopefully future observations, with e.g. HETE-2, will clarify whether
short bursts are associated with host galaxies in the same way as
the long ones.

The pulsar based models have not been studied nearly as extensively as
either the collapsar or merging compact objects, perhaps because the
MHD phenomena they rely on are difficult to simulate numerically. The
magnetic fields and rotation rates invoked for the pulsar models,
though large (P $\sim$ ms; B $\sim$ 10$^{15}$ gauss), are not much
greater than employed for the disk in MHD collapsar models. However,
it is not at all clear how such models would make the large mass of
$^{56}$Ni inferred for SN 1998bw or the highly collimated flow
required to explain energetic events like GRB 990123. Also, the bare
pulsar version of the model\cite{Usov94} ignores the effects of
neutrino-powered winds and the supernova-based version \cite{Wheel99}
ignores the collapse of the massive star that would continue, at least
at some angles, during the few seconds it takes the pulsar to acquire
its large field. A complete calculation of the implosion of the iron
core of a massive star, including the coupled effects of rotation,
magnetic fields, and neutrinos has not been done, but could be in
the next decade.

\section*{Collapsars - Type 1}

We thus consider here a model that can, in principle, satisfy the five
constraints above, at least for long bursts, and has the added virtue
of being calculable, with a few assumptions, on current computers.  A
{\sl collapsar} is a massive star whose iron core has collapsed to a
black hole that is continuing to accrete at a very high rate. The
matter that it accretes, that is the helium and heavy elements outside
the iron core, is further assumed to have sufficient angular momentum
($j \sim 10^{16} - 10^{17}$ cm$^2$ s$^{-1}$) to form a centrifugally
supported disk outside the last stable orbit. The black hole is either
born with, or rapidly acquires a large Kerr parameter. It may also be
possible to create a situation quite similar to a collapsar in the
merger of the helium core of a massive star with a black hole or
neutron star \cite{Fry98}. 

What follows has been discussed in the literature
\cite{Mac99,MWH99,Aloy99}.  The black hole accretes matter along it
rotational axis until the polar density declines
appreciably. Accretion is impeded in the equatorial plane by
rotation. The accretion rate through the disk is insensitive to the
disk viscosity because a steady state is rapidly set up in which
matter falls into the hole at a rate balancing what is provided by
stellar collapse at the outer boundary.  The mass of the accretion
disk is inversely proportional to the disk viscosity and accretion
rates 0.01 - 0.1 M\sun \ s are typical during the first 20 s as the
black hole grows from about 3 M\sun \ to about 4 or 5 M\sun. The
accretion rate may be highly time variable down to intervals as short
as 50 ms \cite{Mac99}, and an appreciable fraction of the matter
passing through the disk is ejected as a powerful ``wind'' that itself
carries up to a few $\times 10^{51}$ erg and a solar mass
\cite{Mac99,Stone99}. Given the high temperature in the disk, this
disk wind will, after some recombination, probably be mostly
$^{56}$Ni. This may be the origin of the light curve of SN 1998bw and
other supernovae associated with GRBs.

Disk accretion also provides an energy source for jets. In the
simplest, but perhaps least efficient version of the collapsar model,
energy is transported from the very hot ($\sim 5$ MeV) inner disk to
the rotational axis by neutrinos. Neutrinos arise from the capture
of electron-positron pairs on nucleons in the disk and deposit a small
fraction of their energy, $\sim$1\%, along the axis where the geometry
is favorable for neutrino annihilation. The efficiency factor for
neutrino energy transport is a sensitive function of the accretion
rate, black hole mass and Kerr parameter, and the disk viscosity
\cite{Pop99}.  Only in cases where the accretion rate exceeds about
0.05 M\sun \ s$^{-1}$ for black hole masses 3 - 5 M\sun \ and disk
viscosities, $\alpha \sim$ 0.1, will neutrino transport be
significant.  Using the actual accretion rate, Kerr
parameter, hole mass as a function of time, and $\alpha \sim 0.1$,
MacFadyen finds for a helium core of 14 M\sun, a total energy
available for jet formation up to $\sim$10$^{52}$ erg. The typical
time scale for the duration of the jet, and a lower bound for the
duration of the GRB, is $\sim$10 s, the dynamical time scale for
the helium core.

In addition to any neutrino energy transport, one has the possibility
of magnetohydrodynamical processes which could, in principle,
efficiently convert a large fraction of the binding energy at the last
stable orbit, up to 42\% $\dot M c^2$, into jet energy. Adopting a
more conservative value, $\epsilon_{\rm MHD} \sim$ 1\% \cite{MWH99},
one still obtains 10$^{52}$ - 10$^{53}$ erg available for jet
formation. Dumping this much energy into the natural funnel-shaped
channel that develops when a rotating star collapses gives rise to a
hydrodynamically collimated jet focused into $\sim$1\% of the sky
\cite{Mac99,MWH99,Aloy99}. Magnetic collimation though uncertain,
could, in principle, increase the collimation factor still further.

Thus jets of equivalent isotropic energy 10$^{54}$, and possibly
10$^{55}$ erg (if, e.g., $\epsilon_{\rm MHD} \sim 0.1$) seem feasible
in this model. The event rate of collapsars is also adequate \cite{FWH99}.

The collapsar model also makes several ``predictions'' some of which
have already been confirmed (these same predictions were inherent in
the original 1993 model \cite{Woo93}. First, the GRB should originate
from massive stars, in fact the {\sl most} massive stars, and be
associated with star forming regions. In fact, given the need for
large helium core mass, collapsars may be favored not only by rapid
star formation, but also by low metallicity. This reduces the loss
of both mass and of angular momentum. Pre-explosive mass loss also provides
a natural explanation for the surrounding medium needed to make the
GRB afterglows and makes a prediction that the density decline as
r$^{-2}$. The GRB duration, $\sim$ 10 s, corresponds to the collapse
time scale of the helium core. The explosion is expected to be highly
collimated, though just how collimated was not realized until 1998
\cite{Mac99}. The jet blows up the star in which it is made so one
expects some kind of supernova. Since the presence of a massive
hydrogen envelope prohibits making a strong GRB, the supernova must be
of Type I (a possible exception would be an extreme Type IIb
supernova, one that had lost all but a trace of hydrogen n its
surface). That the explosion might also produce a lot of $^{56}$Ni
from a disk powered wind was not appreciated until \cite{Mac99}.
Without the $^{56}$Ni, the supernova would have been very dim, which
is why I originally referred to the collapsar model as a ``failed
supernova''.

It also seems natural that both the variable accretion rate
\cite{Mac99} and the hydrodynamical interaction of the jet with the
star which it penetrates may introduce temporal structure into the
burst. Implications for GRB diversity are discussed in $\S$4.

\section*{Collapsars - Type 2}

It is also possible to produce a collapsar in a delayed fashion by
fallback in an otherwise successful supernova \cite{MWH99}. A
spherically symmetric explosion is launched in the usual way by
neutron star formation and neutrino energy transport, but the
supernova shock has inadequate strength to explode the whole
star. Over a period of minutes to hours a variable amount of mass,
$\sim$ 0.1 to 5 M\sun, falls back into the collapsed remnant, often
turning it into a black hole \cite{WW95} and establishing an accretion
disk. The accretion rate, $\sim$0.001 to 0.01 M\sun \ s$^{-1}$, is
inadequate to produce a jet mediated by neutrino annihilation
\cite{Pop99}, but MHD processes may still function with the same
efficiency as in the Type 1 collapsar (or merging neutron stars, for
that matter). Then the total energy depends not on the accretion rate,
but the total mass that reimplodes. For 1 M\sun and $\epsilon_{\rm
MHD}$ = 1\%, this is still 10$^{52}$ erg.

A key difference is the time scale, now typically 10 - 100 times
longer. Thus the most likely outcome of a Type 2 collapsar in a star
that has lost its hydrogen envelope is a less luminous, but longer
lasting GRB. Indeed, there exist GRBs that have lasted hundreds of
seconds and there may be a class of longer, fainter GRBs awaiting
detection. Since black holes may be more frequently produced by fall
back than by failure of the central engine \cite{Fry99}, these sorts
of events might even be more common than ordinary GRBs.

Both kinds of collapsars can also occur in stars that have {\sl not}
lost their envelopes. Stars with lower metallicity have less radiative
mass loss so that solitary stars (or widely detached binaries) might
also end their lives with both a rapidly rotating massive helium and a
hydrogen envelope. Because the motion of the jet head through the star
is sub-relativistic \cite{Aloy99} and because fall back only maintains
a high accretion rate for 100 - 1000 s, highly relativistic jets will
not escape red supergiants with radii $\gtaprx$10$^{13}$ cm. What
happens in more compact blue supergiants is less certain. Generally
speaking, the largest fall back masses will characterize the weakest
supernova explosions and also have the shortest fall back time
scales. With a jet head speed of 10$^{10}$ cm s$^{-1}$, it would have
taken 300 s, for example, to cross the blue progenitor of SN
1987A. The fall back mass in 87A is believed to have been $\ltaprx$0.1
M\sun, probably inadequate to turn the neutron star into a black hole
and certainly too little to make a powerful GRB, but perhaps enough to
make a jet anyway - or at least cause some mixing. Larger mass helium
cores (87A was 6 M\sun) might have more fall back though, definitely
making black holes and more energetic jets. Whether the jet can still
have a large Lorentz factor remains to be calculated.

Even if they do not make GRBs, collapsar powered jets in blue and red
supergiant stars may still lead to very energetic, asymmetric supernova
explosions, possibly accompanied by large $^{56}$Ni production and
luminous soft x-ray transients due to shock breakout \cite{MWH99}.
These transients may have luminosity up to $\sim10^{49}$ erg s$^{-1}$
times the fraction of the sky to which high energy material is ejected
(typically 0.01) and color temperatures of $2 \times 10^6$ K.

\section*{GRB Diversity}

As previously noted, the inferred total energy in gamma-rays for those
GRBs whose distances have been determined is quite diverse.  One
appealing aspect of the collapsar model is that its outcome is
sufficiently variable to explain this diversity. The observed burst
intensity is sensitive not only to the jet's total energy, but also to
the fraction of that energy in the observer's direction that has
Lorentz factor $\Gamma$ above some critical value ($\sim$100). Most
collapsars accrete about the same mass, 1 - 3 M\sun, before accretion
is truncated by the explosion of the star. For an efficiency factor of
1\%, this implies a total jet (and disk wind) energy of $\sim$ few
$\times 10^{52}$ erg. However, depending on the initial collimation of
the jet, its internal energy (or equivalently the ratio of its
pressure to its kinetic energy flux), and its duration, very different
outcomes can result. A poorly collimated jet, or one that loses its
energy source before breaking through the surface of the star may only
eject a little mildly relativistic matter and make, e.g., GRB 980425.
A focused, low entropy jet that lasts $\sim$10 s after it has broken
free of its stellar cocoon might make GRB 990123.

Duration can be affected by such things as the presupernova mass and
angular momentum distribution. Internal energy depends on details of
the jet acceleration. Neutrino powered jets, for example, have much
higher internal energies than some MHD jets and may be harder to
focus. Hydrodynamical focusing of the jet also depends on the density
distribution in the inner disk, which in turn depends on disk
viscosity and accretion rate. And of course the efficiency factor need
not always be 1\%, e.g., for neutrino-powered models and MHD models.

\begin{figure}[h!] 
\centerline{\epsfig{file=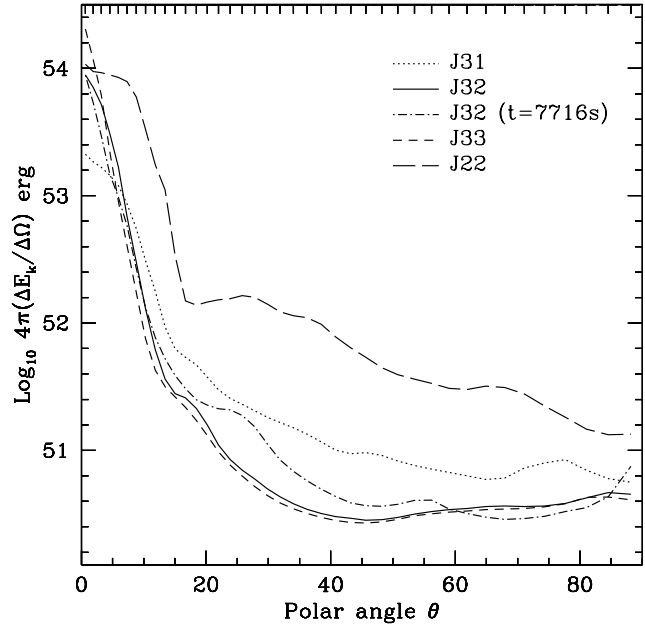,height=3.5in,width=3.5in}}
\vspace{10pt}
\caption{Equivalent isotropic energy as a function of angle for
several models.}
\label{fig1}
\end{figure}

Calculations \cite{MWH99,Aloy99} illustrate this.  Fig. 1 shows the
``equivalent isotropic kinetic energy'' as a function of polar angle
for three models having the same total jet energy, $3 \times 10^{51}$
erg, at the base. All models except the dot-dash line for J22 are
shown 400 s after the initiation of the jet, well after it has broken
out of the helium core. The three models differ only in the ratio of
internal energy to kinetic energy given to the jet at its base. Yet,
even for a constant viewing angle, $\theta = 0$, , the inferred
isotropic energies vary by an order of magnitude. Larger variations
are possible if one goes to other values of viewing angle - {\sl not}
because the GRB is being viewed ``from the side'', but because the
material coming at the observer has both less energy and a lower
Lorentz factor.  Thus it is also possible that GRB 980425 was a more
typical GRB viewed off axis \cite{Nak99}, but not in the sense of a
single highly relativistic beam which emitted a few photons in our
direction. Instead we saw emission from matter coming towards us with
a lower $\Gamma$.

\begin{figure}[h!] 
\centerline{\epsfig{file=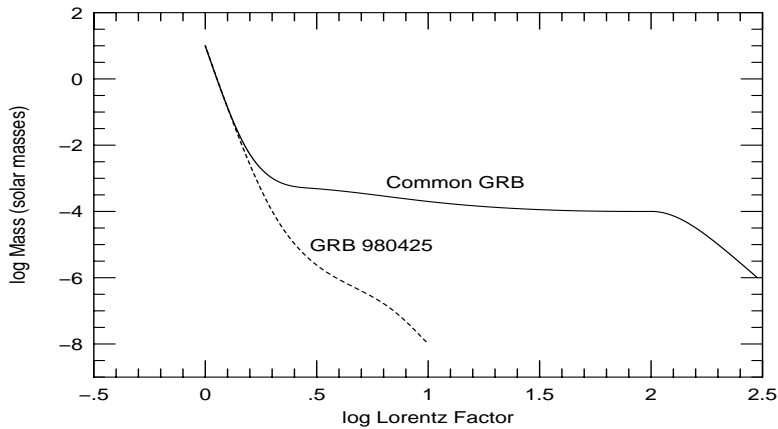,height=3.5in,width=3.5in}}
\vspace{10pt}
\caption{Mass ejected vs. Lorentz factor for two hypothetical models.}
\label{fig2}
\end{figure}

Fig. 2 is not the result of any current calculation, but just a sketch
to illustrate what calculations may ultimately show. (See, for
comparison, Fig. 4 of \cite{Aloy99}, a first pass at one collapsar
model using a code with the necessary relativistic
hydrodynamics. Unfortunately this calculation has not yet been run
long enough to show the final distribution of Lorentz factors).  There
is a large concentration of mass, $\sim$10 M\sun, moving at
sub-relativistic speeds. This is the supernova produced by the jet
passing through the star. Though the speed is ``slow'', most of the
energy may be concentrated here if the jet did not last long enough or
stay focused enough to become highly relativistic (dotted line).  Then
there is a relativistic `tail'' to the ejecta. Even though it is a
small fraction of the mass, this tail could, in some cases, namely the
common GRBs, contain most of the energy in the explosion.

Table 2 indicates some of the diverse outcomes that might arise.  Here
R$_{15}$ is an approximate radius in units of 10$^{15}$ cm where the
material might give up its energy. A typical Wolf-Rayet mass loss rate
has been assumed for those cases where external shocks are clearly
important (x-ray afterglows and GRB 980425). Supernovae also typically
have a photospheric radius of 10$^{15}$ cm. $\Omega/4 \pi$ is the
fraction of the sky into which the mass is beamed. The fractions sum to 
over 100\% because the supernova is not beamed.

\begin{table}[h!]
\caption{Relativistic mass ejected in two artificial models}
\label{table2}
\begin{tabular}{llcrcl}
$\Gamma$ & M/M\sune & E(erg) & $\Omega/4 \pi$ & R$_{15}$(10$^{15}$ cm) & Comment \\
\\
\tableline
 &  &  &  Common GRB  &  &  \\
\tableline
100 & 10$^{-4}$ & 10$^{52}$ & $<1$\% & $<3$ & GRB \\
10  & 10$^{-4}$ & $5 \times 10^{51}$ & 10\% & 3 & X-ray tail\\
1   & 10  &  $5 \times 10^{51}$ & 100\% & 1 & SN Ib/c \\
\tableline
\tableline
 &  &  & GRB 980425  &  &  \\
\tableline
7 & 10$^{-7}$ & 10$^{48}$ & 10\% & 0.01 & GRB 980425 \\
2 & 10$^{-5}$ & 10$^{50}$ & 20\% & 10   & X-ray, radio afterglow \\
1 & 10        & 10$^{52}$ & 100\% & 1    & SN 1998bw \\
\tableline
\end{tabular}
\end{table}

\section*{Time Variability, Lag Time, and Luminosity}

At this meeting we also heard of two fascinating results with
important implications for the use of GRBs as calibrated ``standard
candles'' for cosmology. Ramirez-Ruiz and Fenimore (Paper T-04)
discussed a correlation between ``variability'' and luminosity.  The
more rapidly variable the light curve, the higher the absolute
luminosity. Norris, Marani, \& Bonnell \cite{Norris99} also showed
data to support a high degree of (anti-)correlation between absolute
luminosity and the ``time lag'', the delay time between the arrival of
hard and soft-subpulses. The shorter the lag, the brighter the burst.

Both these effects may be understood as an outcome of Fig. 1.  The
bursts for which we infer the highest luminosities are those that are
observed straight down the axis of the jet, $\theta = 0$. This is also
the angle at which we see the largest Lorentz factors. Slightly away
from $\theta = 0$, both the equivalent isotropic energy and $\Gamma$
drop precipitously.

For larger Lorentz factors, the burst will be produced closer to the
source. Ref. \cite{Pan97} gives a thinning radius where the GRB
becomes optically thin to Thomson scattering that is proportional to
$\Gamma^{-1/2}$. The distance where internal shocks form from two
shells having Lorentz factors $\Gamma_1$ and $\Gamma_2$ is $\Gamma_1
\Gamma_2 c \Delta t$. For smaller radii and larger $\Gamma$, time
scales will thus be contracted. That is larger $\Gamma$ may imply more
time structure on shorter scales and perhaps reduced lag times as well.
Then variability and time lags would be related to the equivalent
isotropic energy because both are functions of the viewing angle.

GRB 980425 is an exception since its GRB was produced by a external
shock interaction between mildly relativistic matter and the
presupernova mass loss.


\begin{references}

\bibitem{Aloy99}
Aloy, M. A., M\"uller, E., Ibanez, J. M., Marti, J. M., \& MacFadyen,
A. I. 1999, submitted to ApJ, astro-ph/9911098

\bibitem{Fry99}
Fryer, C. L. 1999, ApJ, in press, astro-ph/9902315

\bibitem{FWH99}
Fryer, C. L., Woosley, S. E., \& Hartmann, D. H. 1999, ApJ, in press,
astro-ph/9904122

\bibitem{Fry98}
Fryer, C. L., \& Woosley, S. E. 1998, ApJL,  502, L9, astro-ph/9804167

\bibitem{Galama99}
Galama, T. J. 1999, PhD Thesis ``Gamma-Ray Burst Afterglows'', 
Universititeit van Amsterdam

\bibitem{Eberl99}
Janka, H. -Th., Eberl, T., Ruffert, Mm., \& Fryer, C.  1999b, ApJ,
submitted, astro-ph/9908290

\bibitem{Mac99}
MacFadyen, A. I., \& Woosley, S. E. 1999, ApJ, 524, 262,
astro-ph/9810274

\bibitem{MWH99}
MacFadyen, A., Woosley, S. E., \& Heger, A. 1999, submitted to ApJ, 
astro-ph/9910034)

\bibitem{Mes99}
M`esz`aros, P. 1999, Proc. 19th Texas Symposium, astro-ph/9904038, Nuc
Phys B, in press

\bibitem{Nak99}
Nakamura, T. 1999, ApJL, 522, L101

\bibitem{Norris99} 
Norris, J. P., Marani, G. F. \& Bonnell, J. T. 1999, paper G-09 this
meeting, submitted to ApJ, astro-ph/9903233

\bibitem{Pan97} 
Panaitescu, A., Wen, L., Laguna, P., \& M\'esz\'aros, P. 1997, ApJ,
482, 942

\bibitem{Pop99}
Popham, R., Woosley, S. E., \& Fryer, C. 1999, ApJ, 518, 356,
astro-ph/9807028

\bibitem{Ross99}
Rosswog, S., Liebendoerfer, M., Thielemann, F.-K., Davies, M. B.,
Benz, W., Piran, T. 1999, A\&A, in press, astro-ph/9811367

\bibitem{Stone99}
Stone, J. M., Pringle, J. E., Begelman, M. C. 1999, MNRAS, in press,
astro-ph/908185

\bibitem{Usov94}
Usov, V. 1994, MNRAS, 267, 1035

\bibitem{Wheel99}
Wheeler, J. C., Yi, I., H\"oflich, P., \& Wang, L. 1999, ApJ, in
press, astro-ph/9909293

\bibitem{Woo93}
Woosley, S. E. 1993, ApJ, 405, 273

\bibitem{WW95}
Woosley, S. E., \& Weaver, T. A. 1995, ApJS, 101, 181

\end{references}
\end{document}